\documentclass[prl,twocolumn,showpacs,preprintnumbers,superscriptaddress]{revtex4}
\usepackage{graphicx}
\usepackage{color}
\usepackage{dcolumn}
\usepackage{bm}

\usepackage{graphics}

\usepackage{subfig}
\usepackage{amsbsy}
\usepackage{amsfonts}
\usepackage{amsthm}
\usepackage{float}
\newcommand{\ket}[1]{| #1 \rangle}
\newcommand{\bra}[1]{\langle #1 |}

\theoremstyle{plain}

\theoremstyle{definition}

\begin{document}

\title{A Study of Teleportation and Super Dense Coding capacity in Remote Entanglement Distribution}
\author {Sk Sazim, Indranil Chakrabarty}
\affiliation{Institute of Physics, Sainik School Post, 
Bhubaneswar-751005, Orissa, India.\\
 International Institute of Information Technology, Gachibowli, Hyderabad 500 032, Andhra Pradesh, India.}

\begin{abstract}
In this work we consider a quantum network consisting of nodes and entangled states connecting the nodes. 
In every node there is a single player. The players at the intermediate nodes carry out measurements 
to produce an entangled state between the initial and final node.
Here we address the problem that how much classical as well as quantum 
information can be sent from initial node to final node.
In this context, we present strong theorems which state that how the teleportation capability of 
this remotely prepared state is linked up with the fidelities of teleportation of the resource states. Similarly, 
we analyze the super dense coding capacity of this remotely prepared state in terms of the capacities of the resource 
entangled states. 
However, we first obtain the relations involving the amount of entanglement of the resource states with
the final state in terms concurrence. 
These relations are quite similar to the bounds obtained in reference \cite{Gour,Gour1}. 
More specifically, in an arbitrary quantum network when two nodes are not connected, our result
shows how much information, both quantum and classical can be transmitted between these nodes. We show that the amount 
of transferable information depends on the capacities of the inter connecting entangled resources. These results
have a  tremendous future application in the context of determining the optimal path in a quantum network to send the 
maximal possible information. 

\end{abstract}

\maketitle
\section{Introduction}
For a long time, the  concept of entanglement \cite{EPR} was more of philosophical interest rather than a scientific 
discovery. However for the last two decades with the advent of quantum information processing protocols, entanglement is now 
widely recognized as a powerful tool
for implementing tasks that cannot be performed using classical means. It had been seen that quantum entanglement plays a 
pivotal role in large variety  of information
processing tasks such as teleportation \cite{CHB}, super dense coding \cite{CHB2}, remote state 
preparation \cite{AKP} and key generation \cite{NG,MH}.
There are two distinct directions of research involving entanglement. The first one is to find proper criteria of
detecting entanglement while the other one is to have a “good” entanglement measure to quantify the amount of entanglement of a 
given state. For a pure state the von Neumann entropy of the reduced density matrix actually quantifies the 
amount of entanglement present in the system. This is also known as the entanglement entropy of the given entangled 
state. 
Out of various  entanglement measures, the measure concurrence is the  subject of intense research 
\cite{SH,WW,WW1,XG,KC,HF,pran,ZH}. Concurrence was originally derived from the entanglement of formation (EOF) which is used
to compute the amount of entanglement for pure states in two-qubit systems \cite{SH}. Since entanglement of formation is a 
monotonically increasing function of the concurrence, the concurrence itself
can also be regarded as an entanglement measure. A natural way of extending the concept of concurrence for two-qubit mixed 
state is done by convex roof construction \cite{WW}. Afterwards, it had been extended to
arbitrary but finite-dimensional bipartite as well as multiparty systems for both pure and
mixed states \cite{KC,pran}. In an important as well as relevant work, the authors have found that the nature of entanglement present in a multiqubit 
system is monogamous in nature. This physical phenomenon is accredited by a relation involving the concurrences of the 
multiqubit system and its subsystems \cite{kundu}. It was shown that 
concurrence also plays a major in various protocols of RED \cite{Gour,Gour1}.

Teleportation and super dense coding are regarded as the two major achievements of quantum information theory 
\cite{teleport,bennett}. In 
each of these two information processing protocols entanglement plays a pivotal role in execution of them. In short, 
teleportation and super dense coding is all about sending quantum and classical information through a quantum resource. 
In classical information theory we have seen networking plays a key role in sending information from one node to a 
distant node. In quantum networking one of the most challenging problem is to know how much classical and quantum 
information one can send from one node to a distant node which are not initially entangled. In this work we claim to 
provide a solution to this problem by finding out the teleportation fidelity and super dense coding capacity of the
remotely prepared state in terms of teleportation fidelities and super dense 
coding capacities of the  resource states.  In particular, we find out these relations in the context of the entanglement distribution between distant nodes
by the standard swapping of the entangled resource states. But before that we find such relations involving the amount of entanglement of the resource states with
the final state in terms of two different measures of entanglement namely concurrence and entanglement entropy. 
These relations are quite similar 
to the bounds obtained in reference \cite{Gour,Gour1}, where the authors have obtained the bounds in terms of G-concurrence. 
Then by using these relations we establish the relations involving the teleportation fidelity and super dense coding capacity of the 
entangled channels that can be produced in terms of RED protocols.

The organization of the paper is as follows. In section II, we start with two pure entangled resource states 
shared by three parties and obtain the relations involving the  concurrences 
of the  resource states with the final state obtained in the process of RED by swapping. In section III, 
we extend our results  where we have more than two resource states and three parties. 
In section IV,  we provide strong results connecting the teleportation fidelity and super dense coding capability of 
the resource states with the state engineered by the process of entanglement swapping. Finally we 
conclude in section V. 
\section{RELATIONS ON CONCURRENCE FOR TRIPARTITE SITUATION IN REMOTE ENTANGLED DISTRIBUTION (RED)}
In this section, we consider the most simplest situation where Alice and Bob share a pure entangled state $\ket{\psi}_{12}$ 
between them. Similarly, Bob and Charlie also share another entangled state $\ket{\psi}_{23}$ between them. 
This is equivalent of saying, we have entanglement between the nodes $1$ and $2$  as well as between the nodes 
$2$ and $3$. Our aim is to establish the entanglement between the remote nodes $1$ and $3$ which are not initially 
entangled. We adopt the procedure of entanglement swapping to carry out the remote entanglement distribution between
the nodes $1$ and $3$.
In order to swap 
the entanglement, Bob carries out measurement on his qubits which are at the node $2$. 
Interestingly, we find an important relationship between 
the concurrences of the 
entangled states before and after swapping. The most remarkable aspect of this relationship is that this 
tells us about the amount of entanglement that can be created in a remote entanglement distribution (RED) via swapping. 
\subsection{Relations on Concurrence for Two Qubit Pure States}
In this subsection  we start with resource entangled states in $2 \otimes 2$ dimensions. These states are given by
\begin{eqnarray} 
\ket{\psi}_{12}=\displaystyle\sum_{i,j}a_{ij}\ket{ij} \hspace{0.2cm}\mbox{and}\hspace{0.2cm}
\ket{\psi}_{23}=\displaystyle\sum_{p,q}b_{pq}\ket{pq}
\end{eqnarray}
respectively. Here, $a_{ij},b_{pq}\in\cal C$ $(i,j,p,q=0,1)$ are the probability amplitudes satisfying the normalization 
conditions $\sum_{ij}a_{ij}^2=1$ and 
$\sum_{pq}b_{pq}^2=1$.
We consider a situation, where we take into account a general measurement strategy. Here, Bob carries 
out measurement in a non-maximally Bell-type 
entangled basis given by the basis vectors,  $\ket{\phi_G^{rh}}=\frac{1}{\sqrt{B_{rh}}}\displaystyle\sum_{t=0}^1
e^{\pi Irt}R^{rh}_t\ket{t}\ket{t\oplus h}$, where $B_{rh}=\sum_t(R^{rh}_t)^2$ and 
the coefficients $R_j^{rh}$ are defined as
\begin{center} 
\begin{eqnarray}
 R_j^{rh}=\left\{ \begin{array}{rl}
                    n &\mbox{if $(r,h,j)=(0,0,1)$ or $(1,0,0)$},\\
		    m &\mbox{if $(r,h,j)=(0,1,1)$ or $(1,1,0)$},\\
		    1 &\mbox{otherwise}.
                \end{array}\right.
 \end{eqnarray}
\end{center}
Here  the indices 
$n,m(\in\cal C)$ are the entangling parameters and $0\le(n,m)\le1$. And $t\oplus h$ means the sum of $t$ and $h$ modulo $2$.
Now according to general measurements done by Bob on his 
qubits, we have four possible states between the nodes $1$ and $3$ at Alice and Charlie's locations respectively. These four 
possible states based on Bob's measurement outcomes $\ket{\phi_G^{rh}}$ ($r,h=0,1$) are given by,
\begin{equation}
  \ket{\chi^{rh}}_{13}=\frac{1}{\sqrt{M_{rh}}}\displaystyle\sum_{i,q=0}^{1}(\displaystyle\sum_{j=0}^{1}
  e^{-I\pi rj}R_j^{rh}a_{ij}b_{j\oplus h,q})\ket{iq}_{13}.\label{geneq1}
\end{equation}
The modulo sum $j\oplus h$ represents the sum of $j$ and $h$ modulo $2$ and the normalization factors are given by 
$M_{rh}=\displaystyle\sum_{i,q=0}^{1}
(\displaystyle\sum_{j=0}^{1}e^{-I\pi rj}R_j^{rh}a_{ij}b_{j\oplus h,q})^2$. 
Interestingly, here we obtain an important relation between the concurrences of the 
initial and final states,
\begin{equation}
 C(\ket{\chi^{rh}}_{13})=\frac{F_{rh}}{2M_{rh}}C(\ket{\psi}_{12})C(\ket{\psi}_{23}),\label{rel2}
\end{equation}
where the coefficients  $F_{rh}$ are given by,
\begin{center} 
\begin{eqnarray}
F_{rh}=\left\{ \begin{array}{rl}
                    n &\mbox{if $(r,h)=(0,0)$ or $(1,0)$},\\
		    m &\mbox{if $(r,h)=(0,1)$ or $(1,1)$}.
                \end{array}\right.
\end{eqnarray}
\end{center}
This relation (\ref{rel2}) shows that we can always determine the amount of entanglement to be created between the 
unentangled nodes depending upon the choice of the resource states. 
\subsection{Relations on Concurrence for Two Qudit Pure States}
In this subsection we extend our result to the situation where we have entangled states in $d \otimes d$  dimension instead of 
states in $2 \otimes 2$ dimension.
If we know the state properly then we can always rewrite it in the Schmidt decomposed \cite{ESc} form.
If we have a pure 
two-qudit 
state in the form $\ket{\psi}=\displaystyle\sum_{i,j=0}^{d-1}a_{ij}\ket{ij}$ where 
$\displaystyle\sum_{i,j=0}^{d-1}a_{ij}^2=1$, then the Schmidt decomposition form for this state will be
$\ket{\psi}=\displaystyle\sum_{\tilde{i}=0}^{d-1}\lambda_{\tilde{i}}\ket{\tilde{i}\tilde{i}}$, 
where $\displaystyle\sum_{\tilde{i}=0}^{d-1}\lambda_{\tilde{i}}^2=1$ and $\lambda_{\tilde{i}}$ are real and non-negative, 
and {$\{\ket{\tilde{i}}\}$} is an orthonormal basis of the corresponding Hilbert space. The concurrence for two-qudit state $\ket{\psi}$ can be written 
in the form \cite{WW,WW1} 
\begin{equation}
 C(\ket{\psi})=\sqrt{\frac{2d}{d-1}(\displaystyle\sum_{\tilde{i},\tilde{j}=0(\tilde{i}< \tilde{j})}^{d-1}\lambda^2_{\tilde{i}}
\lambda^2_{\tilde{i}})}.\label{simcon}
\end{equation}
For $d=2$, this equation reduces to $C=2\mid\lambda_0\lambda_1\mid$.\\ 

Let us consider a two-qudit pure state shared by parties Alice and Bob
$\ket{\psi}_{12}=\displaystyle\sum_{i=0}^{d-1}\lambda_i\ket{ii}$ and Bob and Charlie shares the pure two-qudit state 
$\ket{\psi}_{23}=\displaystyle\sum_{j=0}^{d-1}\mu_j\ket{jj}$, 
where $\displaystyle\sum_{i=0}^{d-1}\lambda_{i}^2=1=\displaystyle\sum_{j=0}^{d-1}\mu_{j}^2$. In other words, $\ket{\psi}_{12}$ is the 
entanglement shared between the nodes $1$ and $2$, whereas $\ket{\psi}_{23}$ is the entanglement between the nodes 
$2$ and $3$.
Now Bob carries out Bell measurements on his qudits. 
These basis vectors on which the Bell measurements are carried out 
are given by,
\begin{equation}
 \ket{\phi^{rh}}=\frac{1}{\sqrt{d}}\displaystyle\sum_{t=0}^{d-1}e^{\frac{2\pi Irt}{d}}\ket{t}\ket{t\oplus h},
\end{equation}
where $t\oplus h$ means the sum of $t$ and $h$ modulo $d$. The indices $r$ and $h$ can take integer values between $0$ and $d-1$. 
We can revert the above equation to obtain
\begin{equation}
 \ket{ij}=\frac{1}{\sqrt{d}}\displaystyle\sum_{r,h=0}^{d-1}e^{\frac{-2\pi Ijr}{d}}\delta_{i,i\oplus h}\ket{\phi^{rh}}.
\end{equation}
Hence, the combined state of Alice, Bob and Charlie is
\begin{eqnarray}
 \ket{\Phi}_{1223}&=&\ket{\psi}_{12}\otimes\ket{\psi}_{23}\nonumber\\
 &=&\displaystyle\sum_{i=0}^{d-1}\displaystyle\sum_{j=0}^{d-1}\lambda_i\mu_j\ket{ij}_{13}\ket{ij}_{22}\nonumber\\
 &=&\frac{1}{\sqrt{d}}\displaystyle\sum_{i,j}^{d-1}\displaystyle\sum_{r,h}^{d-1}
 e^{\frac{-2\pi Irj}{d}}\lambda_i\mu_j\ket{ij}_{13}\delta_{j,i\oplus h}\ket{\phi^{rh}}_{22}\nonumber\\
 &=&\frac{1}{\sqrt{d}}\displaystyle\sum_{i,r,h}^{d-1}e^{\frac{-2\pi Irj}{d}}\lambda_i\mu_{i\oplus h}
 \ket{i,i\oplus h}_{13}\ket{\phi^{rh}}_{22}.
\end{eqnarray}
According to measurement outcomes $\ket{\phi^{rh}}_{22}$ on Bob's side, the states created between 
the nodes $1$ and $3$  are given by  
\begin{equation}
 \ket{\chi^{rh}}_{13}=\frac{1}{\sqrt{N_{rh}}}\displaystyle\sum_{i=0}^{d-1}e^{\frac{-2\pi Irj}{d}}\lambda_i\mu_{i\oplus h}
 \ket{i,i\oplus h}_{13}\label{qdit11}
\end{equation}
where $N_{rh}=\displaystyle\sum_{i=0}^{d-1}\lambda_i^2\mu_{i\oplus h}^2$ 
are normalization factors. We can construct unitary operators of the form 
$U_{st}=\displaystyle\sum_{r=0}^{d-1}e^{\frac{2\pi isr}{d}}\ket{r}\bra{r\oplus t}$ which can transform states in 
equation (\ref{qdit11}) into its diagonal form.
Hence, the concurrence of the final two-qudit state is given by,
\begin{eqnarray}
 &&C(\ket{\chi^{rh}}_{13})={}\nonumber\\&&\frac{1}{N_{rh}}\sqrt{\frac{2d}{d-1}(\displaystyle\sum_{i<f}^{d-1}
 (\lambda^2_{i}\lambda^2_{f})(\mu^2_{i\oplus h}\mu^2_{f\oplus h}))}.\label{simcon} \label{usol}
\end{eqnarray}
To understand the terms in equation (\ref{usol}) we have to split it in the following way
\begin{eqnarray}
 \displaystyle\sum_{i<f}^{d-1}(\lambda^2_{i}\lambda^2_{f})(\mu^2_{i\oplus h}\mu^2_{f\oplus h})
 =\displaystyle\sum_{i<f}^{d-1}\lambda^2_{i}\lambda^2_{f}
 \displaystyle\sum_{i<f}^{d-1}\mu^2_{i\oplus h}\mu^2_{f\oplus h}\nonumber\\
 -\displaystyle\sum_{i<f}^{d-1}(\lambda^2_{i}\lambda^2_{f}
 \displaystyle\sum_{l< m}^{d-1}\Theta_{lm}^{if}\mu^2_{l\oplus h}\mu^2_{m\oplus h}),
\end{eqnarray}
where function $\Theta_{lm}^{if}$ is defined as
\begin{center}
\begin{eqnarray} 
  \Theta_{lm}^{if}=\left\{ \begin{array}{rl}
                            1 &\mbox{if $(l,m)\neq (i,f)$ for $d\geq 3$},\\
		            0 &\mbox{if $(l,m)=(i,f)$ or $d\leq 2$}.
                             \end{array}\right.
 \end{eqnarray}
\end{center}
It is evident that in case of $d \otimes d$ dimensions, we have no direct relationship as we have obtained in the 
multiqubit case. However we consider a special situation where we have only two non-vanishing Schmidt coefficients, then  
we have the concurrence of the state $|\psi\rangle$ as
\begin{equation}
 C_{ij}(\ket{\psi})=\sqrt{\frac{2d}{d-1}}(\lambda_{i}\lambda_{j}).\label{subcon}
\end{equation}
The $C_{ij}$ are the concurrences of $\ket{\psi}$, when two of the Schmidt coefficients are present only.
Then we have the relation with the concurrences of the initial and final entangled states as
\begin{eqnarray}
 C^2(\ket{\chi^{rh}}_{13})=\frac{(d-1)}{2dN^2_{rh}}[C^2(\ket{\psi}_{12})C^2(\ket{\psi}_{23})-K_d^h]\label{rela123},
\end{eqnarray}
where $K_d^h=\displaystyle\sum_{i<f}^{d-1}(C^2_{if}(\ket{\psi}_{12})
 \displaystyle\sum_{l< m}^{d-1}\Theta_{lm}^{if}C^2_{l\oplus h,m\oplus h}(\ket{\psi}_{23}))$ is a term 
that depends on dimension $d$ and $K_2^q=0$ only when $d=2$. Hence for $d=2$, equation (\ref{rela123}) becomes
\begin{eqnarray}
 C(\ket{\chi^{pq}}_{13})=\frac{1}{2N_{rh}}C(\ket{\psi}_{12})C(\ket{\psi}_{23}).
\end{eqnarray}
This relation involving the concurrences also reflects that the amount of entanglement that can be 
created between the remote nodes is solely a function of the amount of entanglement of the resource 
states. 
\section{RELATIONS ON  CONCURRENCE FOR MULTIPARTY SITUATION IN REMOTE ENTANGLED DISTRIBUTION (RED)}
In this section we study a more general situation where we have more numbers of initial entangled states rather than a pair. 
In other words we are having more than three nodes to begin. The consecutive nodes are entangled. We need to establish 
the entanglement between the initial and the final nodes which are remotely located. We consider the process of entanglement 
swapping as a technique for the remote entanglement distribution (RED). Here in this section we extend these relations obtained 
for entangled qubits in previous section in a more general situation where we have more than three nodes.

Let us 
assume that we have 
$(g+1)$ entangled states in $2 \otimes 2$ dimensions with $g+2$ nodes. In order to obtain an entangled state between 1st and last node we carry out 
$g$ number of entanglement swappings. We separately study two different types of measurement strategies in the entire 
swapping procedure. First of all we consider the case where we carry out simultaneous measurement in a non maximally 
entangled basis in each of these intermediate nodes to obtain an entangled state between the qubits in the first 
and the last node. Secondly, we consider sequential measurements to create successive entanglements between the nodes 
$(1,3)$, $(1,4)$ and finally between the nodes $(1,g+2)$. In each of these cases we obtain the extension of the 
relationships  involving the concurrences of initial and final entangled states.
\subsection{Simultaneous and Sequential Measurement:}
\begin{figure}[t]
\begin{center}

\[
\begin{array}{cc}
\includegraphics[height=2cm,width=7.0cm]{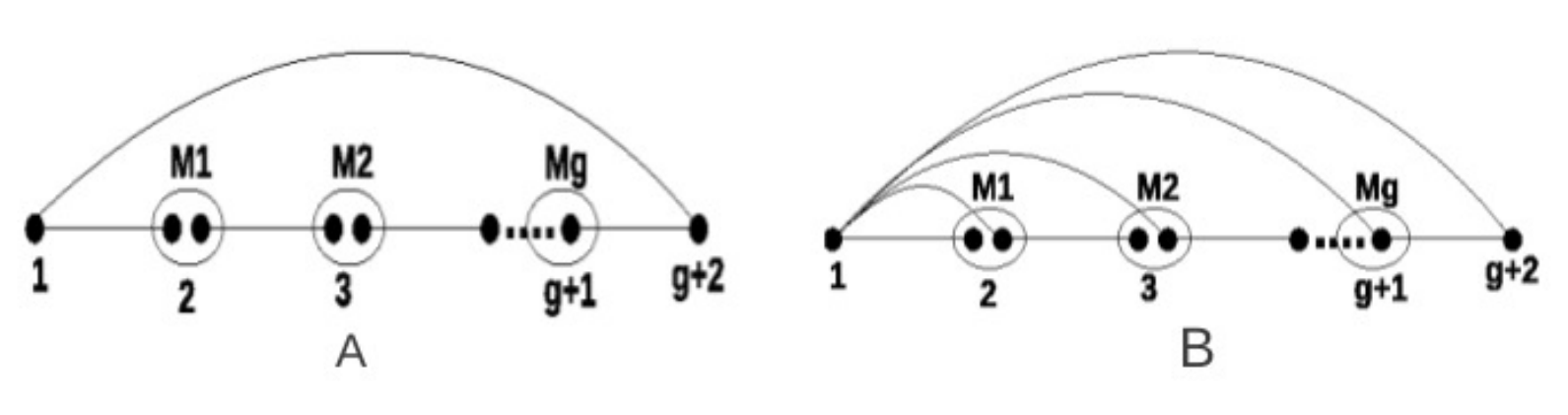}
\end{array}
\]
\caption{In this Figure entanglement swapping is done with simultaneous measurements (A) and sequential measurements (B). 
Here $g$ number of measurements 
$M1, M2,....,Mg$ are carried out  simultaneously (A) and sequentially (B) at the nodes ($2,3,4,...,g+1$) to obtain an 
entangled state between the first and last node (i.e., $1,g+2)$).} \label{simul}
\end{center}
\end{figure}

In this subsection we start with $(g+1)$ entangled states in the most 
general form, $|\psi\rangle=\sum_{i_{k},j_{k}=0}^1a_{i_{k}j_{k}}|i_{k}j_{k}\rangle$, where $k$ denotes the index for the number of 
entangled states and varies from $0$ to $g$. Here for a fixed $k$,  $a_{i_{k}j_{k}}$ denotes the corresponding 
coefficients of the given entangled state. Then we create an entangled state between the qubits at the 
nodes $1$ and $g+2$ by entanglement swapping. In other words we carry out simultaneous measurements $M1, M2,....,Mg$ 
at the nodes $2,3,4,....,g+1$ respectively to obtain an entangled state between the qubits at the 
nodes $1$ and $g+2$ or we carry out measurements $M1, M2,....,Mg$  one after the 
other to create successive entanglement between the pair of nodes $(1,3),(1,4),....,(1,g+2)$ respectively 
[see Fig.(\ref{simul})].

After evaluating the concurrences for the initial states and final state we find them to be related by,
\begin{eqnarray}
 C(\ket{\chi^{r_1h_1,r_2h_2,...,r_gh_g}}_{1(g+2)})=\frac{\displaystyle\Pi_{i=1}^gF_{r_ih_i}}{2^g
 M_{r_1h_1,r_2h_2,...,r_gh_g}}\nonumber\\C(\ket{\psi}_{12})C(\ket{\psi}_{23})........C(\ket{\psi}_{(g+1)(g+2)}).
 \label{relqw}
\end{eqnarray}
Here the indices $r$ and $h$ take the values $0$ and $1$ and the subscript $i(=1,2,...,g)$ denotes the number of 
measurements that have taken place.
The normalization factors are given by $M_{r_1h_1,r_2h_2,...,r_gh_g}=\displaystyle\sum_{i_0,j_{g}=0}^{1}(\displaystyle
\sum_{j_1,j_2,...,j_{g-1}=0}^{1}e^{-I\pi r_1j_0}e^{-I\pi r_2j_1}....e^{-I\pi r_gj_{g-1}}R_{j_0}^{r_1h_1}R_{j_1}^{r_2h_2}$
\\.........$R_{j_{g-1}}^{r_gh_g}a_{i_0j_0}a_{j_0\oplus h_1j_1}a_{j_1\oplus h_2j_2}.......a_{j_{g-1}\oplus h_gj_{g}})^2$.
The superscripts $r_i,h_i\in[0,1], i=1,2,...,g$ comes from the measurement of $g$ parties (i.e., if their measurement results are
$\ket{\phi^{r_1h_1}}\otimes\ket{\phi^{r_2h_2}}\otimes....\otimes\ket{\phi^{r_gh_g}}
=\displaystyle\bigotimes_{i=1}^g\ket{\phi^{r_ih_i}}$). The resultant states obtained after swapping are 
given by,
\begin{eqnarray}
&&\ket{\chi^{r_1h_1,r_2h_2,...,r_gh_g}}_{1(g+2)}={}\nonumber\\&&\frac{1}{\sqrt{M_{r_1h_1,r_2h_2,...,r_gh_g}}}\displaystyle\sum_{i_0,j_g=0}^{1}(\displaystyle
\sum_{j_1,j_2,...,j_{g-1}=0}^{1}e^{-I\pi r_1j_0}e^{-I\pi r_2j_1}{}\nonumber\\&&....e^{-I\pi r_gj_{g-1}}R_{j_0}^{r_1h_1}R_{j_1}^{r_2h_2}
.........R_{j_{g-1}}^{r_gh_g}a_{i_0j_0}a_{j_0\oplus h_1j_1}a_{j_1\oplus h_2j_2}{}\nonumber\\&&.......a_{j_{g-1}\oplus h_gj_{g}})
\ket{i_0,j_{g}}_{1(g+2)}.
\end{eqnarray}
The coefficients, $F_{r_ih_i}$ and $R_{j_i}^{r_ih_i}$ are defined as
\begin{center} 
\begin{eqnarray}
  F_{r_ih_i}=\left\{ \begin{array}{rl}
                    n &\mbox{if $(r_i,h_i)=(0,0)$ or $(1,0)$},\\
		    m &\mbox{if $(r_i,h_i)=(0,1)$ or $(1,1)$}
                \end{array}\right.
\end{eqnarray}  
\end{center}
 and
\begin{center} 
\begin{eqnarray}
  R_{j_i}^{r_ih_i}=\left\{ \begin{array}{rl}
                    n &\mbox{if $(r_i,h_i,j_i)=(0,0,1)$ or $(1,0,0)$},\\
		    m &\mbox{if $(r_i,h_i,j_i)=(0,1,1)$ or $(1,1,0)$},\\
		    1 &\mbox{otherwise}.
                \end{array}\right.
\end{eqnarray}  
\end{center}
\section{RELATIONS ON TELEPORTATION FIDELITY AND SUPER DENSE CODING CAPACITY IN REMOTE ENTANGLED DISTRIBUTION (RED)}
Quantum teleportation and quantum super dense coding are typical information processing
tasks where at present there is intense activity in extending the experimental frontiers [19]. 
In this section we have obtained relation for teleportation fidelity and super dense coding capacity of 
a remotely prepared entangled states with that  of the resource states. These relations are very much important and 
relevant in the context of quantum networking. A quantum network, is a collection of nodes  
interconnected by entangled states that allow sharing of resources and information.
Here we consider quantum network consisting of nodes in sequence and pure entangled states connecting 
consecutive  pair of nodes.
One might ask an important question 
in this context that in an quantum network  what is the amount of quantum information and classical information 
one can send between the initial and final nodes. In other words is there any relation connecting the 
information processing capabilities of the resource states with the final state. We see that indeed there 
is certain relation, which determines 
the amount of information one can send from the 
initial and final node after creating an entangled state between the initial and final node through the process of 
remote entanglement distribution (RED).
In particular we prove a strong theorem which states how the teleportation capability of 
a remotely prepared state is linked up with the fidelity of teleportation of the initial resource states. Similarly, 
we analyzed the super dense coding capacity of the remotely prepared state in terms of the capacity of the initial 
entangled states. In other words these analysis both in case of teleportation and super dense coding shows that the 
amount of information both quantum and classical one can send between two unentangled nodes is dependent on the 
choice of resource states. This actually paves the way in determining the path in an arbitrary quantum network through 
which we can send maximal possible quantum information between any two unentangled nodes.
\subsection{Relations On Teleportation Fidelity In Remote Entangled Distribution (RED)}
Quantum teleportation is all about sending a quantum information of all sending quantum information of one party to 
other with the help of a resource  entangled state.
It is well known that all pure entangled states in $2 \otimes 2$ dimensions are useful for teleportation. However, the situation 
is not so trivial for mixed entangled states.  There arise situation where given an entangled state it can not be 
used as a resource for teleportation. However after suitable local operation and 
classical communication (LOCC) one can always convert an entangled state not useful for teleportation to a state 
useful for teleportation. For a given mixed state,
\begin{eqnarray}
\rho=\frac{1}{4}(I \otimes I +\sum_i r_i.\sigma_i \otimes I+\sum_i s_i.I \otimes \sigma_i 
\nonumber\\+\sum_{ij}t_{ij}\sigma_i \otimes \sigma_j),\label{mix}
\end{eqnarray}
the teleportation fidelity measuring the capability of the state $\rho$ to act as a resource for teleportation, 
is given by,
\begin{eqnarray}
F=\frac{1}{2}[1+\frac{1}{3}(\sum_i \sqrt{u_i})].
\end{eqnarray}
Here in equation (\ref{mix}), $\sigma_i=(\sigma_1,\sigma_2,\sigma_3)$ are the Pauli matrices; $r_i=(r_1,r_2,r_3), s_i=(s_1,s_2,s_3)$ 
are the unit vectors and $t_{ij}$ are the elements of the correlation matrix $T=[t_{ij}]_{3 \times 3}$. The 
quantities $u_i$ are the eigenvalues of the matrix $U=T^{\dagger}T$. A quantum state is said to be useful for 
teleportation when the value of the quantity $F$ is more than the classically achievable limit of fidelity 
of teleportation, which is $\frac{2}{3}$. The entangled Werner state
\cite{wer} in $2 \otimes 2$ dimensions is one example of a useful resource for teleportation for a certain range of classical 
probability of mixing \cite{Horo}. Other examples, of mixed entangled states as a resource for teleportation
are also there \cite{Horo,ind1,ind2}.

In this subsection we investigate how the teleportation fidelities  of the resource states are connected with that 
of the entangled state obtained as a result of entanglement swapping involving the resource states. 
More precisely in a quantum network we consider a situation where we have various resource states 
the consecutive nodes. 
Now, if we want to send the quantum information from the initial node to the final node, one way of doing it is by 
first creating an entangled state between these two nodes, which is obtained after doing measurements on the intermediate nodes 
\cite{Gour, Gour1}. This is known as entanglement swapping and falls into the broader class of remote 
entanglement distribution (RED).  Once the entangled state is created we can send the quantum information.
Here in this subsection we give an important theorems connecting the teleportation fidelities 
(capacities of sending quantum information ) of the resource entangled states with that of the final 
entangled state. This remarkably tells us about the capacity of a quantum network in sending quantum information 
between two desired nodes.

Let us begin with very simplistic situation where there are two parties Alice, Bob  share 
an entangled state $|\psi\rangle_{12}=\displaystyle\sum_{i,j=0}^1 a_{ij}|ij\rangle$ between them, where as Bob and Charlie 
share another state $|\psi\rangle_{23}=\displaystyle\sum_{p,q=0}^1 b_{pq}|pq\rangle$. This is equivalent of saying 
that we have considered  the parties and nodes to be synonymous, then $|\psi\rangle_{12}$ and $|\psi\rangle_{23}$ are the 
respective entangled states between the nodes $(1,2)$ and $(2,3)$. We then propose the following theorem 
connecting the teleportation capability of the resource states with that of the entangled state obtained between the 
non connected nodes as a result of swapping.\\
\textit{\bf Theorem I:}\textit{ For the initial resource states written in the 
form $|\psi\rangle_{12}=\displaystyle\sum_{i,j=0}^1a_{ij}|ij\rangle$ and 
$|\psi\rangle_{23}=\displaystyle\sum_{p,q=0}^1b_{pq}|pq\rangle$, the 
teleportation fidelities of the initial states  and final state $|\chi^{rh}\rangle_{13}$  obtained after 
the measurement in the general basis $\ket{\phi_G^{rh}}=\frac{1}{\sqrt{B_{rh}}}\displaystyle\sum_{t=0}^1
e^{\pi Irt}R^{rh}_t\ket{t}\ket{t\oplus h}$, where $B_{rh}=\sum_t(R^{rh}_t)^2$, 
are related by, $3F(|\chi^{rh}\rangle_{13})-2=
\frac{F_{pq}}{2M_{pq}}\left[3F(|\psi\rangle_{12})-2\right]\left[3F(|\psi\rangle_{23})-2\right]$, where $F_{pq}$ is a 
function of the measurement parameters, $M_{pq}$ are the normalization constants and $(r,h)$ are the indices to 
denote the measurement outcomes.}

\textit{Proof:} We can write fidelities of initial resource states and final remotely 
prepared state as
$F(\ket{\psi}_{12})=\frac{1}{3}(2+C(\ket{\psi}_{12}))$, $F(\ket{\psi}_{23})=\frac{1}{3}(2+C(\ket{\psi}_{23}))$ and 
$F(\ket{\chi^{rh}}_{13})=\frac{1}{3}(2+C(\ket{\chi^{rh}}_{13}))$ respectively,
then just by substituting the values of concurrences in terms of teleportation
fidelities in equation (\ref{rel2}) one can have the relation concerning teleportation fidelities of the initial 
resource states with the final 
remotely prepared state.\\
This gives the more generalized version of the expression relating the teleportation fidelities of thee initial 
resource states with the final remotely prepared states.

Then we consider a complicated situation where we have $(g+1)$ entangled states distributed among hypothetical parties 
in $g+2$ nodes. These entangled states are shared between consecutive nodes. We consider two types of measurement namely 
simultaneous and consecutive measurements $M1, M2,....,Mg$ at $g$ number of nodes [see figure (\ref{simul})]. 
As we have seen in the previous section that both of these measurements create entanglement between the first and 
final node. Here we prove a theorem, quite analogous to previous theorems relating the teleportation capability of 
the resource states with the final state obtained as a result of swapping in the process of remote entanglement distribution 
(RED).

\textit{\bf Theorem II:}\textit{ If we start with $(g+1)$ entangled states in the most 
general form, $|\psi\rangle_{12},|\psi\rangle_{23},....,|\psi\rangle_{(g+1)(g+2)}$, between the nodes $(1,2),(2,3),....(g+1,g+2)$ 
with respective teleportation fidelities $F(|\psi\rangle_{12}),F(|\psi\rangle_{23})....,F(|\psi\rangle_{(g+1)(g+2)})$, then 
the teleportation fidelity of the state $\ket{\chi^{r_1h_1,r_2h_2,...,r_gh_g}}_{1(g+2)}$ is given by
\begin{eqnarray}
&&3F(\ket{\chi^{r_1h_1,r_2h_2,...,r_gh_g}}_{1,(g+2)})-2={}\nonumber\\&&\frac{\displaystyle\prod _{i=1}^{g}F_{r_ih_i}}{2^g
 M_{r_1h_1,r_2h_2,...,r_gh_g}}\left[3F(|\psi\rangle_{12})-2\right]{}\nonumber\\&&
\left[3F(|\psi\rangle_{23})-2\right]...\left[3F(|\psi\rangle_{(g+1)(g+2)})-2\right]\nonumber
\end{eqnarray}}
\textit{Proof:} Just
by substituting the values of the concurrences in terms of the teleportation fidelities in the equation (\ref{relqw}) we 
finally obtain the relation involving the teleportation fidelities of the initial resource states with the final 
remotely prepared state. 

\subsection{Relations On Super Dense Coding Capacity In Remote Entangled Distribution (RED)}
Quantum super dense coding involves in sending of classical information from one sender to the receiver when they are 
sharing a quantum resource in form of an entangled state.
More specifically superdense coding is a technique used in quantum information theory to transmit classical information 
by sending quantum systems \cite{bennett}. In the simplest case, Alice wants to send
Bob a binary number $x \in \{00, 01, 10, 11\}$. She picks up one
of the unitary operators ${I, X, Y, Z}$ according to $x$ she has
chosen and applies the transformation on her multiqubit (the first
multiqubit of the Bell state shared by them). Alice sends her multiqubit
to Bob after one of the local unitaries are applied. The state
obtained by Bob will be one of the four basis vectors, so he
performs the measurement in the Bell basis to obtain two bits
of information. It is quite well known that if we have a maximally entangled state in $H_d\otimes H_d$ as our resource, then we
can send $2 \log_2 d$ bits of classical information. In the asymptotic case, we know one can send $\log_2 d + S(\rho)$ amount of
bit when one considers non-maximally entangled state as resource \cite{agr,hiro,bru,bru1,sha}. It had been seen that the number of classical bits one
can transmit using a non-maximally entangled state in $H_d\otimes H_d$ as a resource is $(1 + p_0\frac{d}{d-1}) \log_2 d$, 
where $p_0$ is the smallest Schmidt coefficient. 
However, when the state is maximally entangled in its subspace then one can send up to
$2*\log_2(d-1)$ bits \cite{ind3}. 
In particular super dense coding capacity for a mixed state $\rho_{AB}$ is defined by
\begin{equation}
 {\cal C}_{AB}=\log_2d+S(\rho_B)-S(\rho_{AB}),
\end{equation}
where, $\rho_B=tr_A[\rho_{AB}]$. Here we note that the expression  $S(\rho_{B})-S(\rho_{AB})$ can either be 
positive or negative. If it is positive then one can use the shared state to transfer bits greater than the classical limit of 
$\log_2d$ bits.
For pure states, $S(\rho_{AB})=0$, then the 
super dense coding capacity is given by,
\begin{equation}
 {\cal C}_{AB}=\log_2d+S(\rho_B)=\log_2d+E(\rho_{AB}),
\end{equation}
where, the entanglement entropy $E(\rho_{AB})$ of the state $\rho_{AB}$ is nothing but the von Neumann entropy 
of the reduced subsystem $\rho_B$.

In this subsection we find how the super dense coding capacities of the resource states are related with the 
super dense coding capacity of the entangled state obtained as a result of entanglement swapping. More precisely in a quantum network we start with various resource states connecting the consecutive nodes, 
and we want to send the classical information from the initial node to the final node. One way of doing it is by 
first creating an entangled states between the end nodes, by doing measurements on the intermediate nodes 
\cite{Gour, Gour1}. Indeed this is the process of entanglement swapping and falls into the broader class of remote 
entanglement distribution (RED).  Once the entangled state is created, we can send the classical information.
Here in this subsection we give an important relationship involving the super dense coding capacities 
(capacities of sending classical information ) of the resource entangled states with that of the final 
entangled state. 

Here we consider only the simplest situation where we have two resource states at our disposal and we want 
to send classical information from one node to another which are not initially entangled.
Let us once again begin with a situation where two parties Alice, Bob  sharing 
an entangled state $|\psi\rangle_{12}=\sum_{i}\lambda_i|ii\rangle$ between them, where as Bob and Charlie 
share another state $|\psi\rangle_{23}=\sum_{j}\mu_j|jj\rangle$ 
( where $\lambda_i, \mu_j$, $(i,j=0,1,...,d)$ are the Schmidt coefficients, satisfying 
$\sum_{i}\lambda_i^2=1$,$\sum_{j}\mu_j^2=1$) with each other \cite{ESc}.  
Then Bob carries out the  Bell state measurement on his qubits at the node $2$ and
according to measurement outcomes $\ket{\phi^{rh}}_{22}$ on Bob's side, the resultant entangled pairs generated 
between the nodes $1$ and $3$ are $\ket{\chi^{rh}}_{13}$ (given in equation (\ref{qdit11})).
The entanglement entropy 
of these states are given by,
\begin{equation}
  E(\ket{\chi^{rh}}_{13})=-\frac{1}{N_{rh}}\sum_i\lambda_i^2\mu_{i\oplus h}^2
  \log_2\left[\frac{\lambda_i^2\mu_{i\oplus h}^2}{N_{rh}}\right].
\end{equation}
Then there arises three situations 
depending upon the choice of the Schmidt coefficients of the resource states. \\

\noindent{\bf Case I:}
First of all we consider the case when both the resource states are maximally entangled i.e., 
when all the Schmidt coefficients are equal to $\frac{1}{\sqrt{d}}$. 
Then the super dense coding capacity of the resource
state is related with the super dense coding capacity of the remotely prepared entangled states $\ket{\chi^{rh}}_{13}$ 
(where $r,h$ are the indices indicating the measurement outcomes ) as,
${\cal C}(\ket{\chi^{rh}}_{13})={\cal C}(|\psi\rangle_{12})={\cal C}(|\psi\rangle_{23})={\cal C}\mbox{(say)}$.
Hence if we have a network consisting of $g+1$ number of maximally entangled states then the super dense coding capacity 
of final state between the nodes $1$ and $g+2$ [see figure (\ref{simul})] will be,
\begin{equation}
 {\cal C}(\ket{\chi^{rh}}_{1,(g+2)})={\cal C},
\end{equation}
where, ${\cal C}={\cal C}(|\psi\rangle_{12})={\cal C}(|\psi\rangle_{23})=...
={\cal C}(|\psi\rangle_{(g+1),(g+2)})$.
 
\noindent{\bf Case II:}
In this particular case we consider the situation when one of the entangled state is maximally entangled 
and the rest is non maximally entangled i.e $\lambda_i=\frac{1}{\sqrt{d}}$, $\mu_j \neq \frac{1}{\sqrt{d}}$, then we have,
${\cal C}(\ket{\chi^{rh}}_{13})={\cal C}(|\psi\rangle_{23})={\cal C}_2\mbox{(say)}$ and if $\lambda_i\neq\frac{1}{\sqrt{d}}$, 
$\mu_j=\frac{1}{\sqrt{d}}$, then we have,
${\cal C}(\ket{\chi^{rh}}_{13})={\cal C}(|\psi\rangle_{12})={\cal C}_1\mbox{(say)}$. Now if we consider a network consisting 
of $g+1$ number of entangled bipartite 
qudit states out of which $n$ number of states are non-maximally entangled and $g+1-n$ number of states are maximally entangled 
then for the strategies in figure (\ref{simul}), the super dense coding capacity of final state 
between the nodes $1$ and $g+2$ will be,
\begin{equation}
 {\cal C}(\ket{\chi^{rh}}_{1,(g+2)})<{\cal C}_p^{\max},
\end{equation}
where, ${\cal C}_p^{\max}$ is the maximum out of $n$ number of super dense coding capacities 
[${\cal C}_p$; $p=1,2,3,....,n$ ]
of non-maximally entangled pure two-qudit resource states.

\noindent{\bf Case III:}
Finally, we consider the case when both the entangled states are not maximally entangled
i.e $\lambda_i\neq\frac{1}{\sqrt{d}}$, $\mu_j \neq \frac{1}{\sqrt{d}}$, then the super dense coding 
capacity of the swapped state is given by, 
${\cal C}(\ket{\chi^{rh}}_{13})<\max[{\cal C}_1,{\cal C}_2]$. And hence easily we can write for a network consisting of $g+1$
number of non-maximally entangled pure two-qudit states, the super dense coding capacity of the final state (as a result of the strategies 
in figure (\ref{simul})) between the nodes $1$ and $g+2$ will be,
\begin{equation}
 {\cal C}(\ket{\chi^{rh}}_{1,(g+2)})<{\cal C}_i^{\max},
\end{equation}
where, ${\cal C}_i^{\max}$ is the maximum out of $g+1$ number of super dense coding capacities 
[${\cal C}_i$; $i=1,2,3,....,(g+1)$ ]
of non-maximally entangled pure two-qudit resource states.
\section{Conclusion}
In a nutshell,  here in this work, we established an important relationship connecting the fidelities of teleportation 
of the resource states with the fidelity of the final state obtained as a result of entanglement swapping.
Similarly we also connected the super dense coding capacities of the resource states with that of the final state.
All these relations are very much important and relevant in the context of quantum networking. These relations 
actually determine the amount of information both classical and quantum, one can send from one node to a desired 
node in a quantum network. In other words, in an arbitrary network when two nodes are not connected, our result
shows how much information both quantum and classical can be sent from one node to other. In fact the amount 
of transferable information depends on the capacities of the inter connecting entangled resources. Depending 
upon the inter connecting entangled resources, we can choose the optimal path in a quantum network to send the 
maximal possible information.
One can investigate the above relations concerning teleportation fidelity in arbitrary dimension in QNet also
\cite{ss}.

\textit{Acknowledgment:} Authors acknowledge Dr. S. Banerjee and Dr. P. Agrawal for having useful discussions, which helped 
in the formulation 
of the central idea of the manuscript. This work is done at Center for Security, Theory and Algorithmic Research (CSTAR), 
IIIT, Hyderabad. Sk Sazim gratefully acknowledge
their hospitality.

\end{document}